# Photo-Nernst Current in Graphene

Helin Cao[1,#], Grant Aivazian[1,#], Zaiyao Fei[1], Jason Ross[2], David H. Cobden[1*], Xiaodong Xu[1, 2,*]

[1]Department of Physics, University of Washington, Seattle, WA 98195, USA
[2]Department of Materials Science and Engineering, University of Washington, Seattle, Washington 98195, USA
[#]These authors contributed equally to this work
[*]Corresponding authors: xuxd@uw.edu, cobden@uw.edu

**Photocurrent measurements provide a powerful means of studying the spatially resolved optoelectronic and electrical properties of a material or device[1-7]. Generally speaking there are two classes of mechanism for photocurrent generation: those involving separation of electrons and holes, and thermoelectric effects driven by electron temperature gradients. Here we introduce a new member in the latter class: the photo-Nernst effect. In graphene devices in a perpendicular magnetic field we observe photocurrent generated uniformly along the free edges, with opposite sign at opposite edges. The signal is antisymmetric in field, shows a peak versus gate voltage at the neutrality point flanked by wings of opposite sign at low fields, and exhibits quantum oscillations at higher fields. These features are all explained by the Nernst effect[8-10] associated with laser-induced electron heating[6,11-14]. This "photo-Nernst" current provides a simple and clear demonstration of the Shockley-Ramo nature of long-range photocurrent generation in a gapless material[5].**

In a semiconductor, electrons and holes generated by photons can live long enough to be spatially separated by an electric field and diffuse to the contacts. This is the basis of photovoltaic cell operation. In contrast, in a gapless material such as graphene the full electron distribution rapidly thermalizes, eliminating the distinction between electrons and holes. Nevertheless, photocurrent is readily produced when light is focused on inhomogeneous regions or junctions in graphene devices[6,7,13-20]. Detailed measurements of the dependence on gate voltage and time delay have shown that it is primarily of a thermoelectric nature[6,11,13,17,19], and that the heating of the electrons is sometimes enhanced by slow energy transfer to the lattice due to the large optical phonon energy and high electron velocity[14,21-26]. However, the means by which a thermoelectric current near the laser spot results in photocurrent in the contacts, which may be located some distance away, has received much less attention.

Unlike in a semiconductor, in graphene one cannot talk about diffusion of majority carriers to the contacts. Rather, in such a gapless material the localized current density source $\boldsymbol{j}_{\mathrm{loc}}$ produces a global photocurrent $I_{\mathrm{ph}}$ by setting up an electric field that drives ambient carriers outside the excitation region and into the contacts, as recently discussed by Song and Levitov[5]. These authors derived an expression giving $I_{\mathrm{ph}}$ as an integral over $\boldsymbol{j}_{\mathrm{loc}}$, analogous to the Shockley-Ramo theorem that gives the current generated between two conducting plates when a charge moves in the insulating space between them[27,28]. Our measurements on graphene devices in a magnetic field demonstrate the existence of the photo-Nernst effect which produces a photocurrent according to this theorem. In the presence of a perpendicular field $\boldsymbol{B}$ a transverse current proportional to $B$ tends to circulate around the laser spot, that is, perpendicular to the electron temperature gradient in the graphene generated by the laser. When the laser is near a free edge, the integral of this chiral $\boldsymbol{j}_{\mathrm{loc}}$ points along the edge, producing a photocurrent which is independent of distance from the contacts. The agreement of the observed dependence on gate voltage and magnetic field with conventional thermoelectric measurements on graphene Hall bars[8-10] firmly establishes the photo-Nernst origin of the response.

We studied two-terminal monolayer devices of graphene on $SiO_2$ selected to have relatively low disorder. Monolayer graphene flakes were prepared by exfoliation onto substrates consisting of 300 nm-thick thermally oxidized $SiO_2$ on highly p-doped silicon, which served as a back gate. Metal (Cr/Au) contacts were patterned by standard e-beam lithography and metallization in an e-beam evaporator. Results from one device are shown throughout, but consistent results were obtained on several devices. An optical image of the device is shown in Fig. 1a. A gate voltage $V_g$ is applied to the conducting Si substrate, and all measurements are made at a stage temperature of 40 K. The two-terminal resistance was measured with an ac 100 nA current excitation at 20 Hz. Fig. 1b shows the conductance at $B = 0$ as a function of gate voltage $\Delta V_g$ measured relative to the conductance minimum (neutrality point), which in this device was at $V_{g0} = 1.3\ V$. The field-effect mobility is roughly $2 \times 10^4$ cm$^2$V$^{-1}$s$^{-1}$. To measure photocurrent, a laser (wavelength 632 nm, spot size ~1 μm, mechanically chopped at 2 kHz) was focused on the surface and the short-circuit in-phase current $I_{ph}$ was detected, in the direction indicated. The laser power was kept at 40 μW, after determining that $I_{ph}$ is linear in power up to 60 μW (Fig. S1). Scanning photocurrent microscopy (SPCM) was performed by measuring $I_{ph}$ as a function of the laser spot position.

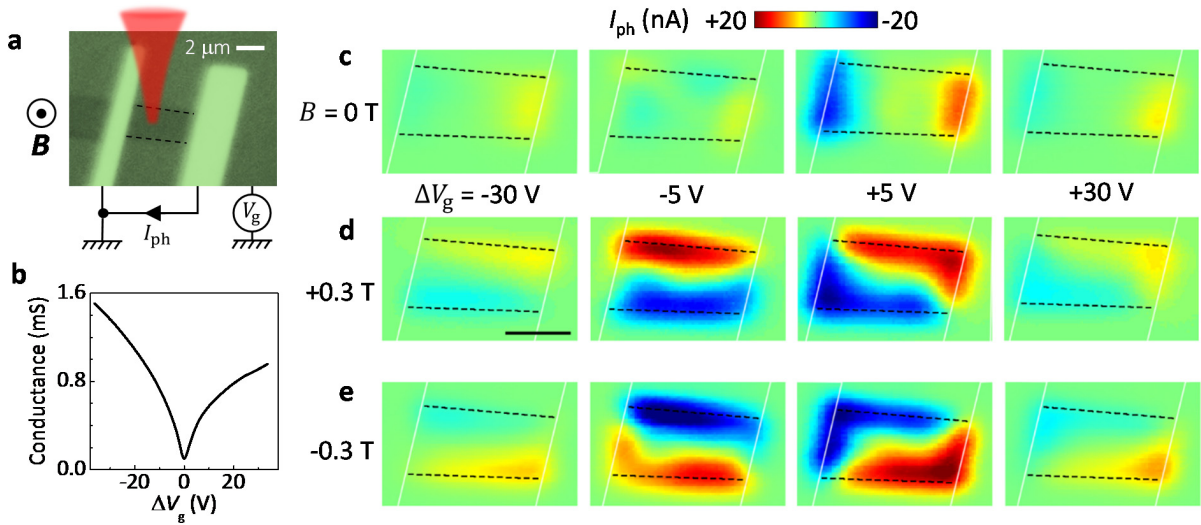

Figure 1 | Scanning photocurrent microscopy (SPCM) of a two-terminal monolayer graphene device in a perpendicular magnetic field. a, Optical image of the device and indication of the experimental setup. The graphene free edges are marked by dashed lines. b, Conductance at 40 K and $B = 0$ vs gate voltage $\Delta V_g = V_g - V_{g0}$ measured relative to the charge neutrality point, $V_{g0} = +1.3\ V$. c-e, SPCM images at $B = 0$ (upper), +0.3 T (middle) and -0.3 T (lower), for the indicated values of $\Delta V_g$. The contact edges are indicated by thin white lines. Scale bar: 2 μm.

Fig. 1c shows SPCM images taken at $B = 0$ at a series of four gate voltages. As usual in higher mobility graphene devices, photocurrent is generated mostly near the contacts and is larger at gate voltages near the neutrality point. Figs. 1d and 1e show corresponding images at $B = +0.3$ T and - 0.3 T. Substantial additional photocurrent is now seen when the laser is near the graphene free edges yet far from the contacts. The polarity of this field-induced photocurrent is opposite at the upper and lower edges and reverses when the field is reversed. In Figs. 2a and b we plot the $B$-symmetric and antisymmetric parts of $I_{ph}$, respectively, at $\Delta V_g = +5$ V. The symmetric part is very similar to the $B = 0$ photocurrent, while the antisymmetric part consists of two stripes, of opposite polarity, parallel to the edges and having almost uniform intensity along each edge.

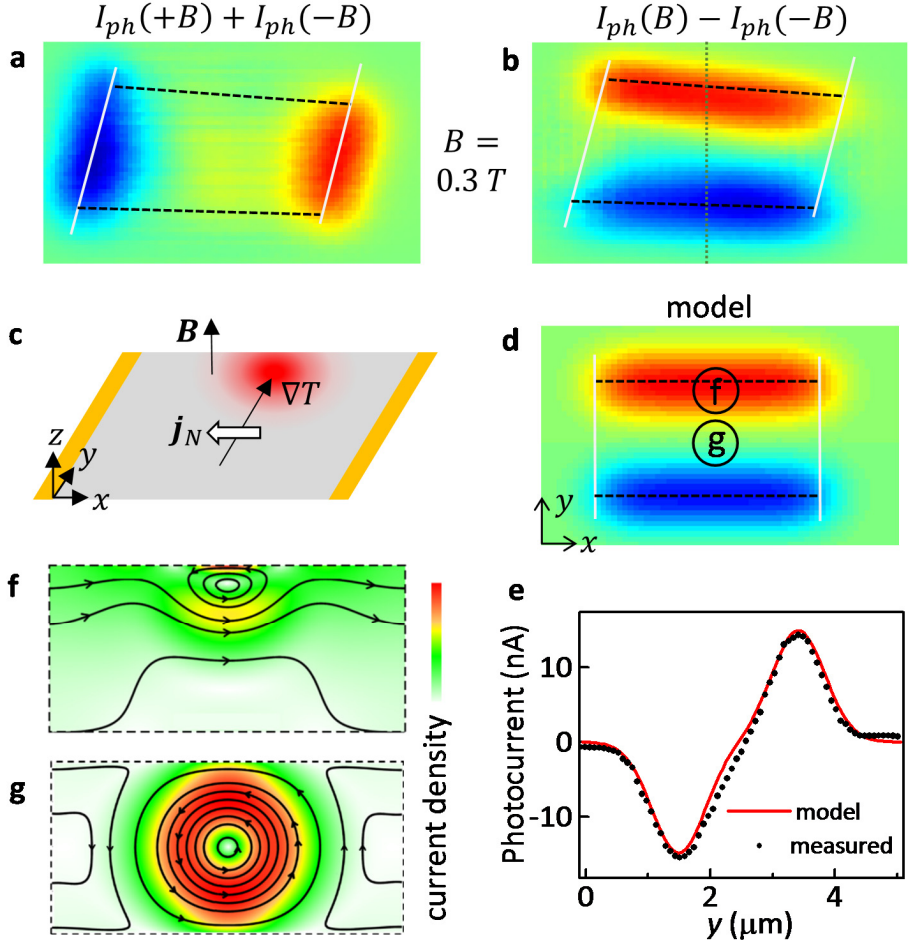

**Figure 2 | Analysis and modeling of the photocurrent induced by a moderate magnetic field. a,** $B$-symmetric and **b,** $B$-antisymmetric components of the photocurrent data shown in Fig. 1 at $\Delta V_g = +5$ V, $B = 0.3$ T. Scale bar: 2 μm. The color scale is the same as in Fig. 1. **c,** Cartoon indicating the electron temperature rise created near the laser spot in a simplified rectangular model of the device, showing the resulting Nernst current $j_N$. **d,** Result of applying Eq. (1) to this model (with a multiplicative factor in the color scale chosen to best match Fig. 2a). **e,** Comparison of model and measured photocurrent along the dotted green line in (b). **f and g,** Calculated $B$-antisymmetric current density profiles within the model, illustrating how photocurrent is generated near the edge (laser at point f) but not in the center (point g).

All of these observations can be understood by starting with the assumption that there is an increase in the electron temperature $T(x, y)$ in and around the laser spot, which generates a thermoelectric current. Since the electron distribution thermalizes in ~100 fs, $T$ should be well defined on a scale of ~100 nm. In similar zero-field measurements it has been shown to rise a few degrees above the lattice temperature[6]. At zero magnetic field and a given gate voltage, the thermoelectric current is parallel and proportional to $\nabla T$, and by mirror symmetry this produces no net photocurrent when the spot is near an edge but far from the contacts. However, in a normal magnetic field $\boldsymbol{B} = B\hat{\boldsymbol{z}}$ there exists a thermoelectric current density component $\boldsymbol{j}_N$ that is odd in $B$ and directed normal to $\nabla T$, as indicated in Fig. 2c. When the spot is near an edge this current density tends to point along the edge. When the spot is near the opposite edge the current density points in the opposite direction. Because of its long-range Shockley-Ramo nature, the resulting photocurrent is independent of the location of the spot along the edge and reflects the directionality of $\boldsymbol{j}_N$. The gate voltage dependence is produced by the dependence of the thermoelectric coefficients on carrier density.

We can model the results quantitatively by applying Song and Levitov's theorem[5] to the extrinsic thermoelectric current $\boldsymbol{j}_{\text{th}} = -\overleftrightarrow{\alpha}\nabla T$, with $\overleftrightarrow{\alpha}$ the 2D thermoelectric tensor. Use of this local relationship is justified since the momentum relaxation length is less than[29] ~1 µm. The total current density $\boldsymbol{j} = \boldsymbol{j}_{\text{th}} + \boldsymbol{j}_{\text{d}}$, which obeys the continuity equation $\nabla \cdot \boldsymbol{j} = 0$, also includes a part $\boldsymbol{j}_{\text{d}}$ (not parallel to $\nabla T$ if $\overleftrightarrow{\alpha}$ is non-diagonal) driven by the electrochemical potential gradient that builds up. The resulting photocurrent is then given by $I_{ph} = \iint \boldsymbol{j}_{\text{th}} \cdot \nabla \psi \, dxdy$. Here the auxiliary scalar field $\psi(x,y)$ is such that $-(\overleftrightarrow{\rho}^T)^{-1}\nabla \psi$ is the current distribution created when a unit bias is applied between the contacts at $\boldsymbol{j}_{\text{th}} = 0$, and $\overleftrightarrow{\rho}$ is the 2D resistivity tensor.

For the particular case of a rectangular strip of isotropic 2D material, with contacts at $x = 0$ and $L$ and free edges at $y = 0$ and $W$, far from the contacts one finds $\nabla \psi = \frac{\beta}{L}(\hat{\boldsymbol{x}} + r\hat{\boldsymbol{y}})$, where $r = \frac{\rho_{xy}}{\rho_{xx}}$ and the coefficient $\beta < 1$ depends on contact details. The photocurrent is then

$$I_{ph} = -\frac{\beta}{L} \iint \left[ \left(\alpha_{xy} + r\,\alpha_{xx}\right)\frac{\partial T}{\partial y} + \left(\alpha_{xx} + r\,\alpha_{xy}\right)\frac{\partial T}{\partial x} \right] dx\,dy \,.$$

The term involving $\frac{\partial T}{\partial x}$, which is even in $B$, gives zero on integration if the temperature rise at the contacts is negligible (or if it is the same at both ends). The other term is odd in $B$ and is proportional to the average temperature difference $\Delta T_{av}$ between the two free edges:

$$-\beta(\alpha_{xy} + r\,\alpha_{xx})\frac{1}{L}\int_0^L dx \int_0^W \frac{\partial T}{\partial y} dy = -\beta(\alpha_{xy} + r\,\alpha_{xx})\Delta T_{av} \,.$$

From the definition of the Nernst coefficient $N$ as the ratio of the transverse electric field to the longitudinal temperature gradient times the magnetic field at zero charge current, it immediately follows that $-(\alpha_{xy} + r\alpha_{xx}) = NB/\rho_{xx}$ for isotropic material. Hence, for a rectangular device, when the laser spot is far from the contacts,

$$I_{ph} = \beta NB\rho_{xx}^{-1}\Delta T_{av} \,. \quad (1)$$

We conclude that the photocurrent is proportional to the Nernst coefficient divided by the longitudinal resistivity. This is likely to remain approximately true for less regularly shaped graphene.

It can be seen from Eq. 1 that this "photo-Nernst" current will be odd in $B$, largest when the laser is near an edge (thus maximizing $\Delta T_{av}$), opposite in polarity at opposite edges, and independent of position along the edge when the laser is far from the contacts. All these facts are consistent with our observations. To be more quantitative, we performed a calculation of $I_{ph}$ for the rectangular strip model. We calculated $\Delta T_{av}$ as a function of laser spot position by solving the heat equation given parameters reported in the literature for graphene on SiO2 of similar mobility (see Supplementary Information). With these parameters, heat flow to the substrate (by optical phonon coupling) dominates lateral heat flow in the graphene[30], unlike in very high mobility devices[31]. In this limit the temperature rises only close to the region where power is absorbed, explaining why photocurrent is produced mostly at distances roughly within the size of the laser spot from the contacts or edges; and $\Delta T_{av}$ is not very sensitive to the thermal conductivity of the graphene. A spatial pattern very similar to that in Fig. 2b was obtained with just two fitting parameters, the laser spot size and an overall multiplicative factor, as shown in Fig. 2d. A comparison of the calculated transverse profile with the measured photocurrent along the dotted green line in Fig. 2b is shown in Fig. 2e.

To provide more intuition for the long-range photocurrent generation process, we show in Figs. 2f and g calculated patterns of the $B$-antisymmetric component of the total current density $\boldsymbol{j}$ that can readily be computed in this simple geometry (see Supplementary Information). When the spot is near the top edge (f), the Nernst effect drives current predominantly to the right (taking $N$ to be negative). Continuity, and the fact that $\boldsymbol{j}_d$ cannot flow in a closed loop, causes the current flow lines to continue into the contacts. When the laser spot is in the center of the strip (g), the net photocurrent is zero, although the Nernst effect produces a circulating current within the graphene. The largest photocurrent occurs slightly inside the edge because the temperature rise is somewhat larger when more of the laser spot is on the graphene.

We now consider the dependence of the photo-Nernst current on $V_g$ and $B$. The Nernst effect in graphene Hall bars has been measured before by conventional means[8-10], and considered theoretically[32]. If the Mott relation[33] applies then $N$ can be written in terms of the components of $\overleftrightarrow{\rho}$:

$$N = \frac{\pi^2 k_B^2 T}{3eB} \frac{\rho_{xx}^2}{\rho_{xx}^2 + \rho_{xy}^2} \frac{\partial}{\partial \epsilon_F}\left(\frac{-\rho_{xy}}{\rho_{xx}}\right).$$

Here $k_B$ is Boltzmann's constant, $e$ the magnitude of the electronic charge, and $\epsilon_F$ the Fermi energy. Putting this into Eq. 1 gives, for a fixed temperature,

$$I_{ph} \propto \frac{\rho_{xx}}{\rho_{xx}^2 + \rho_{xy}^2} \frac{\partial}{\partial \epsilon_F}\left(\frac{-\rho_{xy}}{\rho_{xx}}\right). \quad (2)$$

The Mott relation assumes purely elastic, independent-electron transport and a slowly varying density of states. It should not apply when $\epsilon_F$ is within several $kT$ of the Dirac point in perfect graphene. However, the behavior of transport properties near the neutrality point in real devices is obscured by density inhomogeneities, and deviations from Eq. 2 may be hard to detect.

Fig. 3 shows how the photocurrent depends on gate voltage at a moderate magnetic field, $B = -0.3$ T. Its magnitude and polarity change with $V_g$, while the maxima remain at the same spatial positions near the edges, as can be seen in Fig. 3a. The top axis here shows the 2D carrier density, defined by $n = \frac{C}{e}(V_g - V_{g0})$, taking $\frac{C}{e} = \frac{\epsilon_r \epsilon_0}{e d_{ox}} = 7.2 \times 10^{10}$ cm$^{-2}$/V for oxide thickness $d_{ox} = 300$ nm and $\epsilon_r = 3.9$. $n$ is negative for holes, and loses meaning near the neutrality point due to density inhomogeneities. Fig. 3b shows the dependence on $V_g$ at a fixed laser position (corresponding to the dashed line in Fig. 3a). The photocurrent exhibits a peak at the neutrality point, changes sign for electron or hole density above $\sim 10^{11}$ cm$^{-2}$, and decays slowly at higher densities. Both the peak at the neutrality point and the opposite-sign tails at higher density are consistent with previous Nernst measurements[8-10] and are qualitatively explained by the energy derivative of $\rho_{xy}$ in Eq. 2. If the width of the peak, $\delta n \sim 10^{11}$ cm$^{-2}$, is set by density inhomogeneities, it should be similar to the width of the resistance peak at the neutrality point, shown in Fig. 3c. The two are indeed quite similar. We note that the slight difference in the gate-voltage position of the peaks on the upper and lower edges apparent in Fig. 3a can be explained by a density variation across the sample.

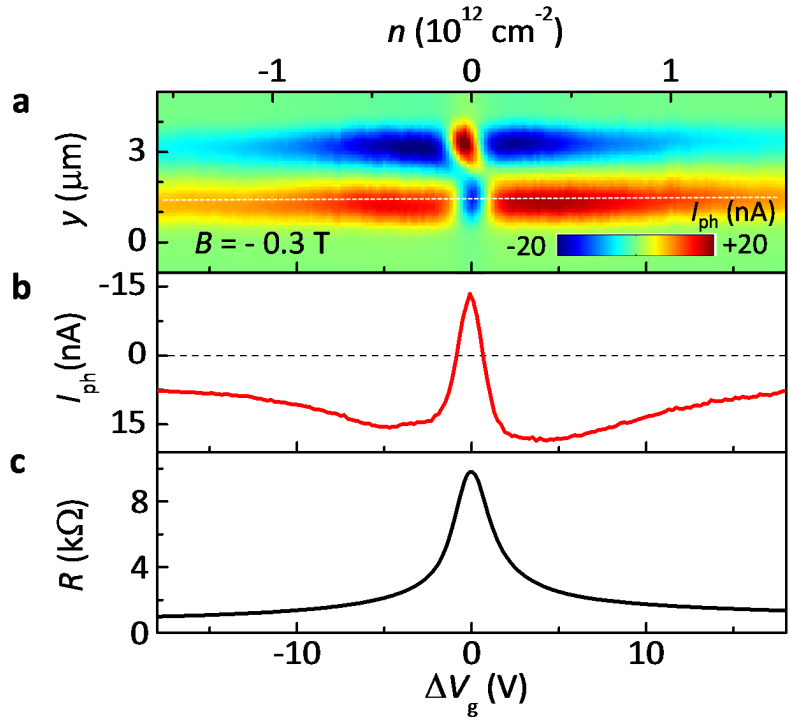

**Figure 3 | Dependence of the photo-Nernst current on gate voltage. a,** Variation of photocurrent with transverse position $y$, along the dotted green line in Fig. 2b, as a function of gate voltage at $B = -0.3$ T. The top axis shows the corresponding 2D carrier density. **b,** Photocurrent vs gate voltage at a fixed position, corresponding to the dashed line in (a). **c,** Two-terminal resistance for comparison.

At higher magnetic field quantum oscillations develop[7] as a function of $V_g$, as illustrated in Fig. 4a. The variation of the peak photocurrent with $V_g$ at a series of magnetic fields is shown in Fig. 4b. For the data at 3 T the filling factor $\nu = nh/eB$ is shown on the top axis. Again, the results resemble previous Nernst measurements[8-10] and are qualitatively explained by Eq. 2, with the oscillations reflecting the energy derivative of the Shubnikov-de Haas oscillations in $\rho_{xx}$. The oscillations are thus large on a small background, in contrast with the much weaker oscillations in the resistance, which is shown in Fig. 4c. We note however that one needs to measure $\rho_{xx}$ employing a multiterminal device to determine $N$ accurately using Eq. (2).

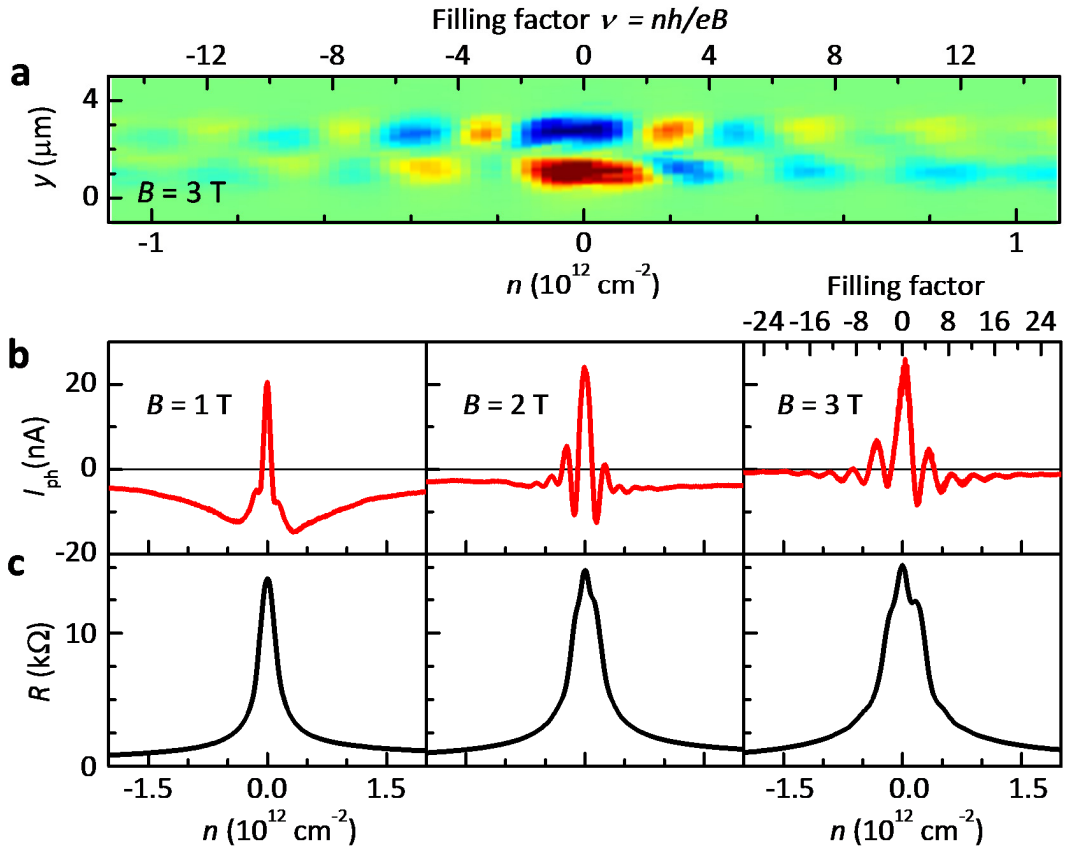

**Figure 4 | Quantum oscillations in the photo-Nernst effect**. **a,** Variation with carrier density of the photocurrent along a line across the graphene strip at 3 T. The top axis shows the corresponding Landau level filling factor. **b,** Maximum photocurrent vs carrier density at a series of magnetic fields. **c,** Two-terminal resistance measured under the same conditions as in (b).

We finish by pointing out that scanning photo-Nernst microscopy could be used as a powerful new probe of materials and devices. For graphene the signal is large and easily detected at low fields (e.g. 50 mT, Fig. S2), and at room temperature (Fig. S3). It can be measured on a simple two-terminal sample, where scanning can reveal spatial variations of properties. It can be separated from other photocurrent contributions by extracting the $B$-antisymmetric signal. Extended to lower temperatures and higher magnetic fields, the technique should yield new insights into the integer and fractional quantum Hall regimes. In those regimes, edge states and correlations may violate the conditions of locality or elastic single-particle transport. Finally, since the photo-Nernst signal shows much stronger quantum oscillations than the resistivity, it can be used to investigate the quasiparticle properties of materials with short scattering times, as has been done with conventional Nernst measurements on cuprates[34] and bismuth[35].

**Acknowledgements**

We thank Boris Spivak, Anton Andreev, Di Xiao, and Chris Laumann for discussions. This work was supported by the National Science Foundation (NSF, DMR-1150719). The experimental setup was partially supported by DoE BES (DE-SC0008145). ZF and DHC are supported by DoE BES (DE-SC0002197). This material is based in part upon work supported by the State of Washington through the University of Washington Clean Energy Institute. Device fabrication was performed at the University of Washington Microfabrication Facility and the NSF-funded Nanotech User Facility.

# Supplementary Information

# Photo-Nernst Current in Graphene

Helin Cao[1,#], Grant Aivazian[1,#], Zaiyao Fei[1], Jason Ross[2], David H. Cobden[1,*], Xiaodong Xu[1,2,*]

[1]Department of Physics, University of Washington, Seattle, WA 98195, USA

[2]Department of Materials Science and Engineering, University of Washington, Seattle, Washington 98195, USA

**Contents:**

Figure S1 | Dependence of photocurrent on laser power
Figure S2 | Photo-Nernst effect at very low magnetic field
Figure S3 | Photo-Nernst effect at room temperature.
S4. Model calculation: photocurrent map
S5. Model calculation: Field-antisymmetric current density map

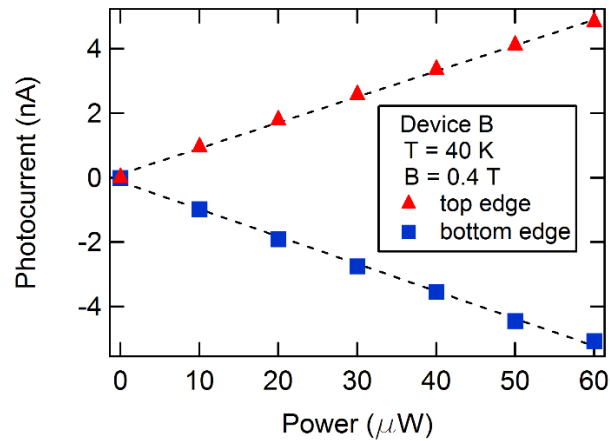

**Figure S1 | Dependence of photocurrent on laser power** for the device used in the main text. The photocurrent $I_{ph}$ was measured at B = 0.4 T with the laser spot near the top (red triangles) and bottom (blue squares) edges of the graphene, at $n = 1.2 \times 10^{12}$ cm$^{-2}$. The black dashed lines are linear fits, showing that the measurements at 40 µW were in the linear response regime.

S1

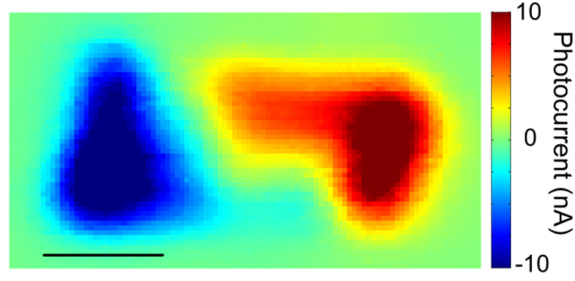

**Figure S2 | Photo-Nernst effect at very low magnetic field.** Scanning photocurrent microscopy (SPCM) image measured at $B = +0.05$ T and $\Delta V_g = 5$ V at $T = 40$ K. The scale bar is 2 μm.

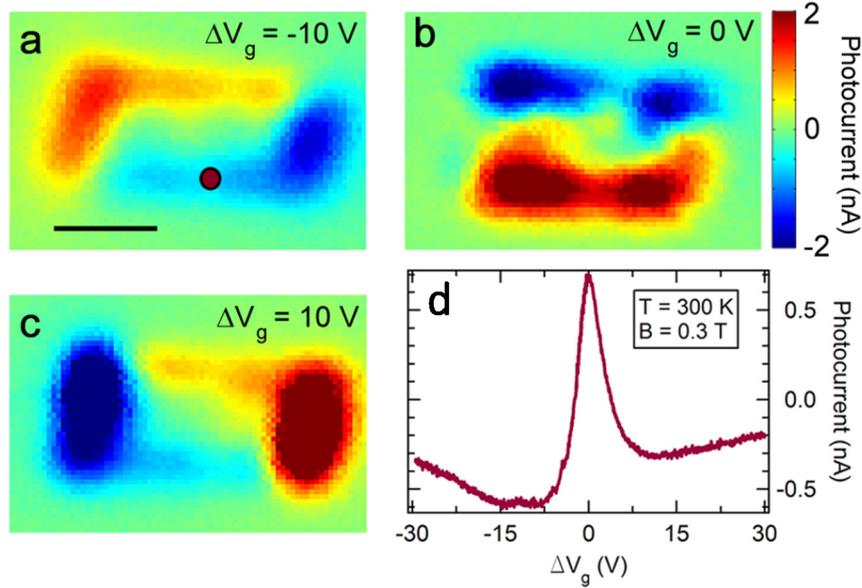

**Figure S3 | Photo-Nernst effect at room temperature.** (a)-(c), Room-temperature SCPM images at $B = +0.3$ T for different gate voltages. (d) Gate dependence of the photocurrent measured with the laser spot at the brown dot in (a). The scale bar is 2 μm.

## S4. Model calculation: photocurrent map

We calculated the photocurrent as a function of laser position using Eq. (1) in a model with a rectangular channel defined by $0 < x < L$, $0 < y < W$. We took the laser spot to have a Gaussian shape, and modeled the temperature profile using the following partial differential equation for the temperature increase $T(x, y)$ in the graphene due to laser heating:

$$\kappa D \nabla^2 T - GT = -\gamma P_0 \exp\left(-\frac{r^2}{r_0^2}\right).$$

Here, $\kappa = 100$ W/K·m is the in-plane thermal conductivity of graphene on SiO$_2$[1,2]; $D = 0.35$ nm is the thickness of graphene; $G = 1000$ W/K·cm$^2$ is the thermal coupling between graphene and the SiO$_2$ substrate[3]; $\gamma$ is the absorption coefficient; $P_0$ is the peak intensity; $r$ is distance from the center of the laser spot; and $r_0$ is the beam radius. The values of $\kappa$ and $G$ used were those at 40 K to match the experiment. The metal electrodes with edges at $x = 0, L$ were taken to be thermal reservoirs with fixed temperature ($T = 0$), while the free edges, at $y = 0, W$, were thermally isolated ($\partial T/\partial y = 0$).

Given $T(x, y)$, the laser-generated current density is

$$\boldsymbol{j_{ph}} = -\vec{\alpha}\nabla T = -\alpha_{xy}\left(\frac{\partial T}{\partial y}, -\frac{\partial T}{\partial x}\right) - \alpha_{xx}\left(\frac{\partial T}{\partial x}, \frac{\partial T}{\partial y}\right) = \alpha_{xy}\hat{z} \times \nabla T - \alpha_{xx}\nabla T,$$

using $\alpha_{xx} = \alpha_{yy}$ and $\alpha_{xy} = -\alpha_{yx}$. From this we calculate $I_{ph}$ employing Song and Levitov's result.[4] For a strip of graphene of length $L$, far from the contacts the weighting field is

$$\nabla\psi(r) = \frac{\beta}{L}\left(\hat{x} + \frac{\rho_{xy}}{\rho_{xx}}\hat{y}\right).$$

The coefficient $\beta < 1$ includes effects of contact resistance. Near the contacts $\nabla\psi(r)$ deviates from this form in a complicated way.

$$I_{ph} = \int \boldsymbol{j_{ph}}(r) \cdot \nabla\psi(r) d^2r = -\frac{\beta}{L}\left(\alpha_{xy} + \frac{\rho_{xy}}{\rho_{xx}}\alpha_{yy}\right)\int_0^L \int_0^W \frac{\partial T}{\partial y}dxdy$$

This expression can be simplified in two ways. First, the intergral is related to the average temperature difference between the two edges:

$$\frac{1}{L}\int_0^L \int_0^W \frac{\partial T}{\partial y}dxdy = \frac{1}{L}\int_0^L [T(x, 0) - T(x, W)]\, dx = \Delta T_{av}$$

Second, the coefficient in brackets is proportional to the the Nernst coefficient $N$. This relates the transverse electric field to the longitudinal temperature gradient, $E_y = NB\frac{\partial T}{\partial x}$, when the electric current is zero, $\boldsymbol{j} = \sigma \boldsymbol{E} - \alpha\nabla T = 0$, i.e., $\boldsymbol{E} = \rho\alpha\nabla T$. Assuming isotropy in the plane,

$$NB = (\rho\alpha)_{yx} = \rho_{yy}\alpha_{yx} + \rho_{yx}\alpha_{xx} = \rho_{xx}(-\alpha_{xy}) + (-\rho_{xy})\alpha_{xx},$$

and so

$$\left(\alpha_{xy} + \frac{\rho_{xy}}{\rho_{xx}}\alpha_{xx}\right) = -\frac{NB}{\rho_{xx}}.$$

Thus we obtain Eq. 1 in the main text,

$$I_{ph} = \beta NB\rho_{xx}^{-1}\Delta T_{av} = C\Delta T_{av},$$

where the coefficient $C$ is independent of laser position. To match the predicted photocurrent profile (Figure 2d) with the measured profile (Figure 2b), we calculated $\Delta T_{av}$ as a function of laser position for different values of $r_0$, and then chose $r_0$ and $C$ to optimize the match. The result was $r_0 = 0.6$ µm, which is close to the beam spot size. With the above parameters the temperature rise does not extend far outside the laser spot because of heat flow into the substrate.

## S5. Model calculation: Field-antisymmetric current density map in Figs. 2f, g

The current density maps shown in Figs. 2f and 2g were obtained as follows. The total current density is

$$\boldsymbol{j} = \boldsymbol{j}_{ph} + \boldsymbol{j}_d,$$

where $\boldsymbol{j}_d = -\sigma \nabla \phi$ is the diffusion current density and $\phi$ the electrochemical potential. To find $\boldsymbol{j}(x,y)$ we thus need to find $\phi(x,y)$. In the steady state $\boldsymbol{j}$ is divergence free:

$$\nabla \cdot \boldsymbol{j} = \nabla \cdot (\boldsymbol{j}_{ph} + \boldsymbol{j}_d) = 0,$$

$$\therefore \nabla \cdot (\boldsymbol{j}_d) = -\nabla \cdot (\boldsymbol{j}_{ph}) = -\nabla \cdot (\alpha_{xy} \hat{\boldsymbol{z}} \times \nabla T - \alpha_{xx} \nabla T),$$

$$\therefore \nabla \cdot (\sigma \nabla \phi) = \alpha_{xx} \nabla^2 T.$$

Now we write $\phi = \phi_S + \phi_A$ where $\phi_S$ is even and $\phi_A$ is odd in $B$. We consider only small $B$, so to first order $\phi_S$, $\alpha_{xx}$ and $\sigma_{xx}$ are independent of $B$ while $\phi_A$, $\alpha_{xy}$ and $\sigma_{xy}$ are proportional to $B$. We then have

$$\sigma_{xx} \nabla \phi_S = \alpha_{xx} \nabla T \quad (1)$$

and

$$\nabla^2 \phi_A = 0.$$

We can therefore obtain $\phi_A$ by solving Laplace's equation numerically with appropriate boundary conditions. At the edges of the metal electrodes ($x = 0, L$) we have $\phi_A = 0$ because $\phi = 0$. At the free edges ($y = 0, W$), $j_y = 0$, and so

$$j_{d,y} = -j_{ph,y},$$

$$\therefore \sigma_{xy} \frac{\partial \phi}{\partial x} - \sigma_{xx} \frac{\partial \phi}{\partial y} = -\alpha_{xy} \frac{\partial T}{\partial x} + \alpha_{xx} \frac{\partial T}{\partial y}.$$

The $B$-antisymmetric part of the last equation is

$$\sigma_{xy} \frac{\partial \phi_S}{\partial x} - \sigma_{xx} \frac{\partial \phi_A}{\partial y} = \alpha_{xy} \frac{\partial T}{\partial x},$$

but from Eq. 1 above, $\sigma_{xx} \frac{\partial \phi_S}{\partial x} = \alpha_{xx} \frac{\partial T}{\partial x}$, and $\sigma_{xx} = \rho_{xx}^{-1}$ to first order in $B$, so the boundary condition on $\phi_A$ at $y = 0, W$ is

$$\frac{\partial \phi_A}{\partial y} = \frac{1}{\sigma_{xx}} \left( \frac{\sigma_{xy}}{\sigma_{xx}} \alpha_{xx} + \alpha_{xy} \right) \frac{\partial T}{\partial x} = N \frac{\partial T}{\partial x},$$

where $N$ is the Nernst coefficient. After finding $\phi_A$ as described above, the part $\boldsymbol{j}_A$ of the current density proportional to $B$ was constructed from

$$\boldsymbol{j}_A = \alpha_{xy}\hat{\boldsymbol{z}} \times \nabla T - \sigma_{xx}\nabla\phi_A .$$

The flow lines of $\boldsymbol{j}_A$ and the mapping of intensity to color in Fig. 2 were selected judiciously for illustrative purposes.

**Supplementary References**